\documentclass[preprint,proceedings]{rmaa}

\usepackage{paralist}

\usepackage{psfrag,color}

\SetYear{2002}
\SetConfTitle{Energetics of Cosmic Plasmas}

\title{Statistical Synthesis Models} 

\author{
  M. Cervi\~no,\altaffilmark{1} 
  V. Luridiana\altaffilmark{1}}

\altaffiltext{1}{Instituto de Astrof\'\i sica de Andaluc\'\i a (CSIC),
Granada, Spain.}

\shortauthor{Cervi\~no \& Luridiana}
\shorttitle{Statistical Synthesis Models}

\fulladdresses{
\item Miguel Cervi\~no: Instituto de Astrof\'\i sica de Andaluc\'\i a
(CSIC), Camino Bajo de Hu\'etor 24, Apartado 3004,
Granada 18008, Spain (\email{mcs@laeff.esa.es}).
\item Valentina Luridiana: Instituto de Astrof\'\i sica de Andaluc\'\i a
(CSIC), Camino Bajo de Hu\'etor 24, Apartado 3004,
Granada 18008, Spain (\email{vale@iaa.es}).

}

\listofauthors{M. Cervi\~no \& V. Luridiana}
\indexauthor{Cervi\~no, M.}
\indexauthor{Luridiana, V.}

\abstract{ In this contribution we evaluate the minimal cluster mass for
which the effects of an incomplete sampling of the Initial Mass Function
(IMF) cannot be neglected.  This minimal cluster mass corresponds to the
situation in which the integrated luminosity of the cluster modeled equals
the luminosity of the most luminous individual star included in the model,
and it takes values between $3 \times 10^2$ and $6 \times 10^5$ M$_\odot$
depending on the age and the observed band. We show different examples for
young ($t < 10$ Myr) and old ($t$ up to 10 Gyr) stellar populations.  We
also make here a first release of the spectral energy distribution (SED)
from 160 $\mu$m ($1.8 \times 10^{12}$ Hz) to 25 keV ($6.0 \times 10^{18}$
Hz) for star-forming regions with ages between 0.1 and 10 Myr, and
metallicities between $Z$=0.001 and $Z$=0.040. The SEDs are available at our
WWW server and include the corresponding quantities for the evaluation of
sampling effects.  These SEDs can be directly used to obtain colors
(including the X-rays ones), as input of photoionization codes, or for
$\chi^2$ fitting with observed data taking into account the intrinsic
uncertainty of the models due to the IMF sampling.}

\resumen{En este trabajo evaluamos la masa m\'\i nima de c\'umulos
estelares, ${\cal M}^{min}$, para la cual los efectos de muestreo de la
Funci\'on Inicial de Masa (FIM) no pueden ignorarse. Esta ${\cal M}^{min}$
corresponde a la situaci\'on para la cual la luminosidad del c\'umulo es
igual a la luminosidad de la estrella individual m\'as brillante incluida
en el modelo, y toma valores entre $3 \times 10^2$ y $6 \times 10^5$
M$_\odot$ seg\'un la edad y la banda observada. Mostramos ejemplos para
poblaciones j\'ovenes ($t < 10$ Ma) y viejas ($t$ hasta 10 Ga). Tambi\'en
hacemos p\'ublica la distribuci\'on espectral de energ\'\i a (DEE) desde
160 $\mu$m ($1.8 \times 10^{12}$ Hz) hasta 25 keV ($6.0 \times 10^{18}$ Hz)
para regiones de formaci\'on estelar con edades entre 0.1 y 10 Ma y
metalicidades entre $Z$=0.001 y $Z$=0.040. Las DEEs est\'an accesibles en
nuestro servidor WWW e incluyen las cantidades necesarias para evaluar los
efectos de muestreo de la FIM. Las DEEs pueden usarse tanto para la
obtenci\'on de colores (incluyendo los de rayos X), como para entrada de
modelos de fotoionizaci\'on o para ajustes $\chi^2$ de datos observados
teniendo en cuenta la incertidumbre intr\'\i nseca en los modelos debida al
muestreo de la FIM.}

\addkeyword{Galaxies: Star clusters}
\addkeyword{Galaxies: Stellar content}
\addkeyword{Open Clusters and Associations: General}

\begin{document}
\maketitle

\section{The need for a statistical modeling}
\label{sec:intro}

Evolutionary synthesis models have been extensively used since the work of
Tinsley \& Gunn~(1976). Unfortunately, the limitations for their use have
not been extensively studied. In this work we present the most basic limit
for the use of standard synthesis models (those that assume a continuously
populated Initial Mass Function, IMF). A more detailed study can be found
in Cervi\~no \& Luridiana 2003 (CL03, submitted).

Maybe, the most trivial limit for the usage of a synthesis model is the
following one:
{\it The total luminosity of the cluster modeled must be larger than the
individual contribution of any of the stars included in the model.  }
This obvious statement defines a natural theoretical limit that has not
always been considered when synthesis models are applied to real
observations.  Based on this limitation, we can establish a {\it Lowest
Luminosity Limit} (LLL) for the application of synthesis models, which
corresponds to {\it the situation where the integrated luminosity of the
cluster modeled equals the luminosity of the most luminous individual star
included in the model}.

Since the integrated luminosity scales with the initial amount of gas
transformed into stars, ${\cal M}$, a minimal initial amount of gas
transformed into stars, ${\cal M}^{min}$, can be inferred for the use of
synthesis models.  In clusters with masses below ${\cal M}^{min}$, sampling
effects in the IMF cannot be neglected and a statistical modeling is
needed. Even more, it can be demonstrated that even the results of
synthesis models for clusters with ${\cal M} < 10 \times {\cal M}^{min}$
have an intrinsic relative uncertainty equal or larger than 10\%.

\section{Some examples of the LLL}

\begin{figure}[!t]
  \includegraphics[width=\columnwidth]{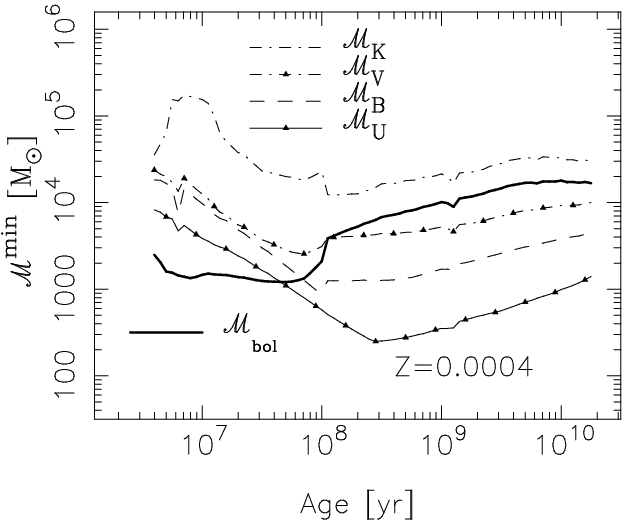} \caption{${\cal
  M}^{min}$ as a function of the age for different photometric bands.}
  \label{fig:Mcolor}
\end{figure}

In order to illustrate how relevant this effect may be, we show in
Fig.\@ \ref{fig:Mcolor} the ${\cal M}^{min}$ values at different ages and
photometric bands. In this case the ${\cal M}^{min}$ values have been
computed from the isochrones by Girardi et al.\@ (2002) with metallicity
$Z$=0.0004, and simple stellar populations (or Instantaneous Burst, IB)
results, and it is assumed a Kroupa~(2001) IMF in the mass range 0.01 -- 120
M$_\odot$. 

\begin{figure}[!t]
  \includegraphics[width=\columnwidth]{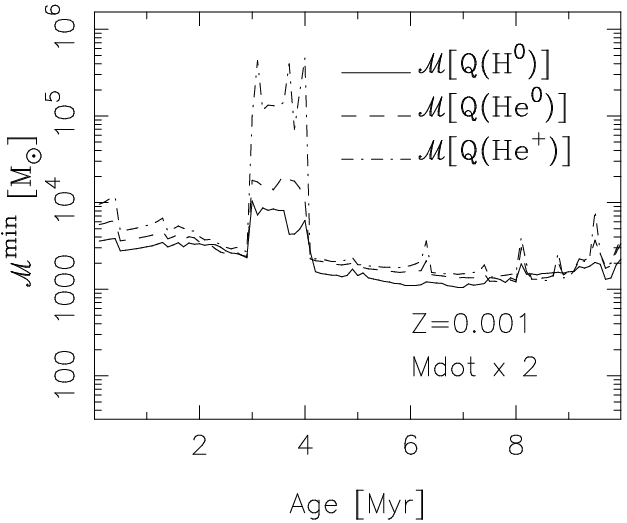} \caption{${\cal M}^{min}$
  as a function of the age for ionizing photons above the H$^0$, He$^0$ and
  He$^+$ edges.}
  \label{fig:Mion}
\end{figure}

In the case of young stellar populations (ages lower than 10~Myr) we show
in Fig.\@ \ref{fig:Mion} the ${\cal M}^{min}$ values corresponding to the
number of ionizing photons above the H$^0$, He$^0$ and He$^+$ ionization
edges. The ${\cal M}^{min}$ values correspond to models following a
Salpeter (1995) IMF in the mass range 0.1 -- 120 M$_\odot$ and they have
been computed from a hybrid version of the codes by Cervi\~no, Mas-Hesse \&
Kunth\@ (2002, CMHK02) and Leitherer et al.\@(1999) in order to both
compute ${\cal M}^{min}$ and include the atmosphere models from Smith,
Norris, \& Crowther~(2002). In the figure we use evolutionary tracks with
high mass-loss rates from Meynet et al.\@ (1994). X-ray emission has not
been considered in this figure.

Note that, in the galactic context, the most massive OB association known,
Cygnus OB2 (c.f. Kn\"odlseder 2000), has transformed into stars $\sim 10^5$
M$_\odot$ of gas. Hence, the modeling of {\it normal} galactic and
extragalactic \ion{H}{ii} regions, with a lower amount of gas transformed
into stars, is potentially affected by sampling effects.

\section{Statistical Synthesis Models}

As we have seen before, the IMF sampling may play an important role in the
determination of the evolutionary status of stellar clusters obtained via
synthesis models. Some studies from a qualitative point of view can be
found in, e.g., Cervi\~no, Luridiana, \& Castander (2000), Bruzual (2002),
and references therein. However, a quantitative theoretical formalism is
needed to address this subject and apply the results to real data.

Examples of such kind of formalism can be found in Lan{\c{c}}on \& Mouhcine
(2000), or in Cervi\~no et al.\@ (2002). These last authors, based on the
definition by Buzzoni (1989) of an effective number of stars, $\cal N$,
evaluate quantitatively the sampling effects in stellar clusters.  The
formalism is completely valid for any age range and for quantities that
scale linearly with $\cal M$. It also gives a first order estimation of the
bias in the ratios or logarithmic quantities predicted by synthesis models
when the sampling effects are quite important (Cervi\~no \& Valls-Gabaud
2003), or, equivalently, when the observed cluster have $\cal M$ values
close to the LLL (see CL03 for more details).

Now, we make here a first public release of the complete spectral energy
distribution (SED) and the corresponding $\cal N$ values at each wavelength
needed to evaluate the sampling effects.  The results come from an improved
and extended version of the synthesis code presented in Cervi\~no \&
Mas-Hesse (1994), and have been discussed in CMHK02. A first experimental
public release of the code will be available in the following months.

The SEDs range from 160 $\mu$m ($1.8 \times 10^{12}$ Hz) to 25 keV ($6.0
\times 10^{18}$ Hz), for ages between 0.1 and 10 Myr and for metallicities
between $Z$=0.001 and $Z$=0.040.  They can be directly used to obtain
colors (including the X-rays ones), as input of photoionization codes, or
for $\chi^2$ fitting with observed data taking into account the intrinsic
uncertainty of the models due to the IMF sampling.  The data, together with
the sampling effects for other observables, can be
obtained in tabular form at:

\begin{center}
{\tt http://www.laeff.esa.es/users/mcs/SED/}.
\end{center}

\begin{figure}[!t]
  \includegraphics[width=\columnwidth]{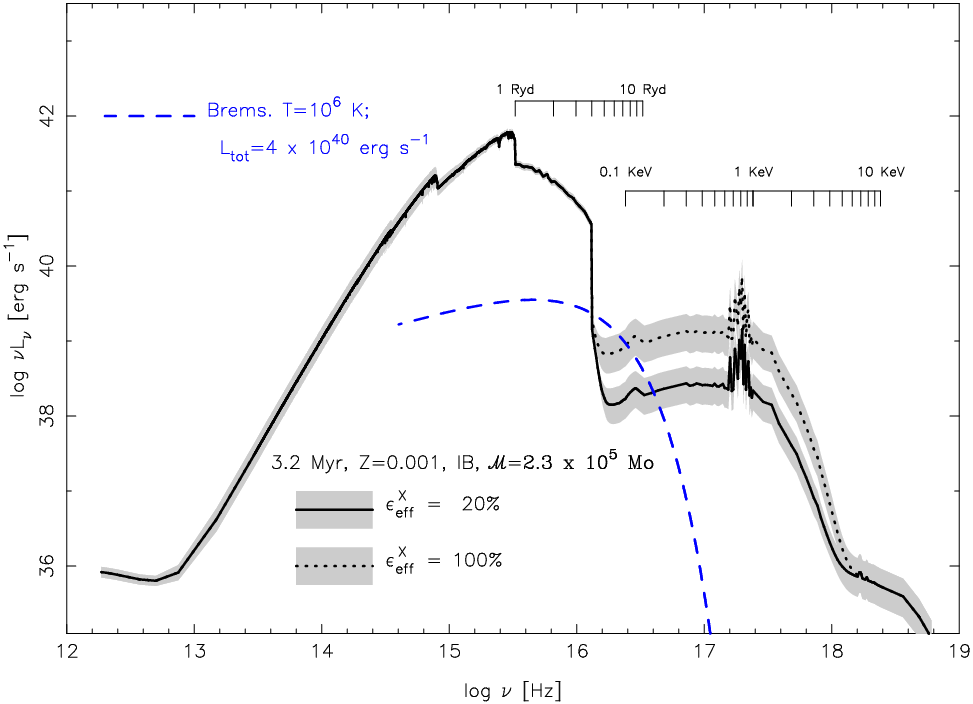}
  \caption{Multiwavelength emission for a 3.2 Myr star-forming region with
  the corresponding 90\% Confidence Limits. A bremsstrahlung component with
  $10^6$ K and a total luminosity of $4\times 10^{40}$ erg s$^{-1}$ is also
  shown for comparison.}  \label{fig:X}
\end{figure}

One of the most important features of our newly realesed SEDs is the
extension in the X-ray domain.  As an example of a possible application of
this feature, we considered the work by Stasi{\'n}ska \& Izotov (2003), who
study the conditions needed to reproduce the observed sequences of low
metallicity star-forming regions, address the difficulties that traditional
models (the ones without X-rays) would find to reproduce at the same time
{\it all} the observational constraints, and point out directions where one
should go in order to find physical solutions to the problems encountered.
Among other effects, they show that an additional high-energy component is
needed to explain the observed trend in diagnostic diagrams, and, as an
example, they use a bremsstrahlung spectrum with $10^6$ K and a total
luminosity of $4\times 10^{40}$ erg s$^{-1}$ (Fig.\@ \ref{fig:X}, dashed
line). Incidentally, this component corresponds actually to the type of
{\it supersoft} X-ray sources observed by {\sc rosat} (Rappaport et al.\@
1994).  In Fig.\@ \ref{fig:X} we show the multiwavelength emission
predicted by our code for a 3.2 Myr star-forming region with $Z$=0.001 and
${\cal M}$=2.3$\times 10^5$ M$_\odot$ (i.e., ${\cal M}$=10$^5$ M$_\odot$ in
the mass range 0.8 -- 120 M$_\odot$, as in Stasi{\'n}ska \& Izotov 2003)
for different efficiencies of conversion of kinetic energy in X-rays.  The
kinetic energy included in our model is produced by stellar winds and
supernovae (CMHK02), so it is totally self-consistent with the stellar
population responsible for the ionization of the gas. The comparison
between the bremsstrahlung component by Stasi{\'n}ska \& Izotov (2003) and
the high-energy portion of our spectrum shows that the availability of SEDs
extending into the high-energy range might help to explain some observed
features of stellar populations; more specifically, the newly released SEDs
will give the opportunity to perform a more detailed analysis of the
effects of X-rays in the optical emission line spectrum.

\acknowledgements

We thank Grazyna Stasi{\' n}ska for positive feedback. MC has been
partially supported by the AYA 3939-C03-01 program.  VL is supported by a
Marie Curie Fellowship of the European Community programme {\it Improving
Human Research Potential and the Socio-Economic Knowledge Base} under
contract number HPMF-CT-2000-00949.

\end{document}